\documentclass[journal]{journal}
\usepackage{algorithm} 
\usepackage{algpseudocode} 
\usepackage{amsmath,amsthm,amssymb,graphicx,hyperref}

\ifCLASSINFOpdf
\else
\fi
\begin{document}
%
\title{An Authentic Algorithm for Ciphering and Deciphering Called  \textit{Latin Djokovic} }
%
%
%

\author{Diogen~Babuc
\thanks{Diogen Babuc is with the Computer Science Department, West University of Timișoara, Bd. Vasile Pârvan 4, Timișoara, Romania.}}

\maketitle

\begin{abstract}
The question that is a motivation of writing is how many devote themselves to discovering something in the world of science where much is discerned and revealed, but at the same time, much is unknown. The insightful elements of this algorithm are the ciphering and deciphering algorithms of Playfair, Caesar, and Vigenère. Only a few of their main properties are taken and modified, with the aim of forming a specific functionality of the algorithm called \textit{Latin Djokovic}. Specifically, a string is entered as input data. A key $k$ is given, with a random value between the values $a$ and $b=a+3$. The obtained value is stored in a variable with the aim of being constant during the run of the algorithm. In correlation to the given key, the string is divided into several groups of substrings, and each substring has a length of $k$ characters. The next step involves encoding each substring from the list of existing substrings. Encoding is performed using the basis of \textit{Caesar algorithm}, i.e. shifting with $k$ characters. However, that $k$ is incremented by $1$ when moving to the next substring in that list. When the value of $k$ becomes greater than $b+1$, it will return to its initial value. The algorithm is executed, following the same procedure, until the last substring in the list is traversed. Using this polyalphabetic method, ciphering and deciphering of strings are achieved. The algorithm also works for a 100-character string. The $x$ character isn't used when the number of characters in a substring is incompatible with the expected length. The algorithm is simple to implement, but it's questionable if it works better than the other methods, from the point of view of execution time and storage space. 

\end{abstract}


%
\IEEEpeerreviewmaketitle

\section{Introduction}
%
%
%
%
\IEEEPARstart{L}{iving} in an era where many things are going on intensively, some individuals are replicating what they have been provided. However, there are also individuals, many indeed, who produce something authentic, conforming to what is already present; while the exceptional ones create something starting from an unstable base.
How many individuals dedicate themselves to exploring the realm of science, where many things are discovered and unveiled, but also where much remains unknown is a question that intrigues us. Based on an essential parameter from Maslov's hierarchy of human needs, the author has built, assembled with insightful elements, the following: an algorithm of ciphering and deciphering. 

The secure communication must be processed, and it includes an encoding process and a decoding process at the end of sending, respectively at the last reception of the communication system.

Sufficient security solution details are present for various system functionalities and capabilities. These include secure commerce and payments through private communications, as well as password protection. 

The main scope of the ciphering sphere is to hide some significant information or observation from a base. It involves a procedure of understanding and using mathematical properties to encode and decode data. Doing so allows us to store sensitive information or transmit it over unsecured networks, such as the Internet so that it cannot be read by anyone except the intended recipient. Even the creators of the database where the information is stored should not access the data. The process of determining the encoded message and information hiding is in close relationship with the information theory, engineering, and various security systems \cite{IEEEhowto:h}. 

The step-by-step traversal of an algorithm, in a detailed manner, is controlled by both the given specifications and a key \cite{IEEEhowto:j}. This is seen as the secret parameter for a particular messaging context. Keys are important because ciphers without variable keys are facile to crack and less useful. 

\hfill October 2022

\subsection{General analysis}
It is well known that \textit{Caesar algorithm} works by following the procedure of shifting the character by $k$ positions \cite{IEEEhowto:i}. After the entire text has been traversed, all characters in that string must be shifted. 

What has been collected from the \textit{Playfair algorithm} is to split the string into several substrings, each one having the length of $k$. In the algorithm proposed by the author, the initial value of $k$ is considered to perform the division of the initial string. As for \textit{Vigenère algorithm}, the increment or modification of the key $k$ has been contemplated. This implies that the algorithm is a polyalphabetic one. 

To take each algorithm separately, the Caesar algorithm is typical and simple to understand \cite{IEEEhowto:i}. It changes the letter from the alphabet with a new letter that is shifted by $k$ positions. It is not stable in the modern era, in the age of computers and technology. Using the brute force method it can be broken easily. This happens because only $25$ distinct key variations are possible.

The Vigenère algorithm, compared to the Caesar algorithm, provides some level of security with the introduction of a keyword. This keyword is executed again to traverse the entire text. The Vigenère table contains the letters of the alphabet in the form of rows and columns. Using the table, the algorithm transforms each letter and we receive an encoded message \cite{IEEEhowto:g}. 

The Playfair algorithm is another example of a classic cipher that has a $5\times5$ square with alphabetic letters arranged in a suitable manner \cite{IEEEhowto:c}. We can select a key and place it in the array. The remaining letters of the English alphabet are then placed one by one into the cipher matrix. If the letter is already present in the array due to the key, then it will not be added again. Next, the string is split into pairs and if a pair has the same character, then they are separated by inserting a padding letter like $x$. For the others, if the pair are different alphabetic letters and are in the same array row, then each letter is modified with the letter before it \cite{IEEEhowto:b}. 

\subsection{Related literature}
In this section, methods similar to the chosen topic will be analyzed. The main objectives of these works and the advantages and disadvantages will be stated.

The main goal of the next paper \cite{IEEEhowto:i} is to develop a variant of the algorithm that combines, and uses, Caesar cipher and Vigenère cipher that can prevent unauthorized access or modification of sensitive information, which should remain private. Classic ciphers cannot boast of this. The paper aims at the mixed realization of the Caesar cipher and Vigenère cipher with a 95\% confusion relation. 

It misses introducing a new and secure scramble in two classic digits so as to produce a strong algorithm that is not too facile to crack by brute force or frequency analysis.
It is desired to create an improved cipher for a better-to-perform plaintext cipher. The code that performs the same procedures at each step is modified because it is easy to access, the logic not being a deep one \cite{IEEEhowto:a}. 

If an input is changed by a little, the output changes significantly. This will make cracking the cipher very difficult if a step in the algorithm is omitted. Data security will be a good one.

An altered version of the Vigenère algorithm was proposed in the \textit{Enhancing Security of Vigenere Cipher by Stream Cipher} work by Ali and Sarhan, which presents the process of adding a random bit to each byte before the message is encoded. This technique fails in the procedure to find the length of the key. The main disadvantage of this method/technique is that the size of the encoded message will be increased by about less than 60\%.

A new approach was presented by combining the Vigenère encoding key with the $LFSR$ (Linear Feedback Shift Register) key \cite{IEEEhowto:g}. The proposed technique generates a series of keys with pseudorandom letters. In the usual Vigenère cipher, each alphabet has a fixed numerical value, but in the present technology, we have eight tables. In each table, each alphabet letter represents a different numerical value. It can create a problem for the receiver to read the message by inserting spaces between words, and the receiver has to guess the exact place to insert the space in the decoded plaintext.

A method specified in the work \cite{IEEEhowto:g} proposed some variations on the traditional Vigenère cipher. First, the key must be any type of character, for example, math symbols, numbers, and punctuation instead of characters. Second, an arbitrary number is introduced for the key in order to spread the options so that only specialists are able to understand the message. Another method is mentioned in the paperwork \cite{IEEEhowto:g}, which permits extending the original Vigenère table to 92 characters. It includes over 60 additional characters from the original table and redesigns the character mapping structure. Although it includes a large character set and introduces case sensitivity in the encoding/decoding process, it cannot perform space, single quote, and backslash encoding because they are not part of the table.  So it's better, but not ideal. 

Another work disclosed and illustrated an algorithm that uses the prime number, its primitive roots, and the quotient that generates them to give some advantages to the Vigenère cipher and the Caesar cipher \cite{IEEEhowto:i}.
This algorithm can accept plaintext containing letters, upper and lower case letters, numbers, and special characters. 

Thus, the user can effortlessly encode the combination of letters, numbers, and characters in a timely manner. If a clear message is given, the $5\times5$ rules will apply, followed by some modifications. When the plaintext character that is in the same pair as its partner occurs again, the first character is modified with the character to the right, with the first element of the line circularly following the last. The second character is modified with the character on the left, with the last element of the line circularly following the first. If a word consists of an odd number of characters, it will add the $None$ character to complete the pairs, since the $None$ character cannot affect the clear text for the decoding stage.

Here the characters are not linearly changed but are randomly shifted by using the substitution box, \textit{s-box}, and permutation technique which is implemented in modern techniques like \textit{Blowfish method} \cite{IEEEhowto:h}. Then, the substitution box must be created by implementing the technique of similar figures. 
The characters are then changed with the equivalent values that refer to the substitution box. In addition, they scramble the ciphertext to hide language features using permutation techniques (randomly changing the positions of characters in the ciphertext). Ciphertext permutation is done when using double-column transposition on the ciphertext. The proposed algorithm can encode a series of clear texts that the Caesar or Vigenère algorithm does not have the ability to encode.

The modified Playfair algorithm has some advantages in using password structure and inclusion \cite{IEEEhowto:f}. First, we can embed spaces in the decoded plaintext. Second, using arrays, we can store information about whether an $X$ appeared, because of the same alphabet appearing in that chart, or whether it appeared because of an odd number of alphabets in the plaintext \cite{IEEEhowto:e}. 

Finally, the sender will also need to provide a password that will be encoded due to the key table, and this encoded password will be attached to the encoded message and sent. At the other end, the receiver will have to provide the password and the key, if the password matches. Then, only the subsequent decoding of the message is performed, otherwise it will not apply.

However, the algorithm faces some drawbacks, such as that a text cannot be encoded for itself. Second, there is reciprocity between plaintext and ciphertext, unless they are in the same row or column. Third, for any chart, there can be close to $700$ combinations regardless of key. Fourth, the original message cannot be recovered from the ciphertext in many cases. If we use the Playfair $16\times16$ algorithm, then the key length can be bigger than $25$ characters \cite{IEEEhowto:d}. 

Vigenère cipher only encodes alphabet letters. A plaintext message containing uppercase and lowercase alphabets must be converted to uppercase before encoding. There is no case sensitivity in the previous algorithm. In addition, the algorithm does not provide the ability to encode numbers, spaces, special characters, or symbols that are used in the English language. The researchers proposed an extended version of the Vigenère cipher using a $95\times95$ Vigenère table, which provides the ability to encode lowercase letters, numbers, and 33 special characters \cite{IEEEhowto:g}, including the space that is used on the keyboard. The proposed algorithm is equivalent to the unique algorithm because it uses modulo 95 instead of using modulo 26.

\subsection{Concrete methods}
The significant components of the technique are the sequences of the encoding/decoding algorithms: Playfair, Caesar, and Vigenère. The aim is to create a specific functionality of the algorithm called \textit{Latin Djokovic} by taking and modifying some of the main properties of the previously mentioned algorithms. The input data consists of a string that is entered. A value of $a$, \{$a<48$\} will be read at the first step; that means $b$ could be at most $50$. Given a key $k$, its value is selected randomly from the range between $a$ and $b$, where $b=a+3$, with the mention that $k$ can be at most $51$ (lowercase letters of the English alphabet + uppercase letters - 1). The obtained value of $k$ is stored in a variable, denoted with $k_{init}$, with the aim of being constant during the algorithm's run. The string is divided into multiple groups based on the provided key (similar to the \textit{Playfair algorithm}, though with a few different specifications) where each string has a length of $k$ characters; the exception may be the last substring, in case the size of the initial string is not divisible by $k$. 

\begin{algorithm}
\caption{\textit{Latin Djokovic}: Division }\label{alg:caption}
\begin{algorithmic}
\Require $ls \gets $ vector of substrings  
\Function{divideInGroups} {$str$, $k$}
\For {$i \gets 0,\ str.length()-1,\ i+k$}

    \For {$j \gets i,\ k+i$}
    \State $substring \gets substring + str_j$
    \EndFor
    \State Add $substring$ to a vector of substrings
\EndFor
\State return vector of substrings
\EndFunction
\end{algorithmic}
\end{algorithm}

\noindent The text is divided into several subgroups of length $k$, by concatenating character by character. This method will return an array of substrings for encoding or decoding. 

The subsequent step involves encoding each substring present in the list.

\begin{algorithm}
\caption{\textit{Latin Djokovic}: Ciphering }\label{alg:caption}
\begin{algorithmic}
\Require $str \gets $ encoded string 
\State $str \gets $ initial string 
\State $a \gets value_1$
\State $b \gets a+3 $
\State $k \gets value_2$
\State $k_{init} \gets k $
\State $list\_substrings \gets$ \Call {divideInGroups}{$str$, $k$}
\For {$string$ in $list\_substrings$}
\State \Call {cipheringUsingCaesarAlg}{$string$, $k$}
\State $k \gets k+1$
\If{$b+1 < k $}
    \State $k \gets k_{init} $
\EndIf
\EndFor
\end{algorithmic}
\end{algorithm}

The encoding process involves utilizing the Caesar algorithm as the basis, which involves shifting the characters by a specific number of positions, precisely by $k$ characters. But, as the list of substrings is traversed, the value of $k$ is increased by $1$ for each subsequent string. When $k$ becomes greater than $b+1$, It will return to its original value. 
The method is performed by following the same process until the last substring in the list is ciphered.
The same principle is followed in the string decoding part, the only difference is the shifting part, which will work in reverse (shifting to the left), using \textit{Caesar algorithm}. The key $k$ will increment each time we traverse the list, from one substring to another. 

\begin{algorithm}
\caption{\textit{Latin Djokovic}: Deciphering }\label{alg:caption}
\begin{algorithmic}
\Require $str_e \gets $ decoded/clear string 
\State $str_e \gets $ encoded string 
\State $a_1 \gets a$
\State $b_1 \gets b $
\State $k \gets k_{init}$
\State $list\_substrings \gets$ \Call {divideInGroups}{$str_e$, $k$}
\For {$string_e$ in $list\_substrings$}
\State \Call {decipheringUsingCaesarAlg}{$string_e$, $k$}
\State $k \gets k+1$
\If{$b+1 < k $}
    \State $k \gets k_{init} $
\EndIf
\EndFor
\end{algorithmic}
\end{algorithm}

It will go through each element in the list, each encoded substring, and decode it by using \textit{Caesar algorithm}. The value for $k$ will increase until it reaches $b+2$. Then, $k$ will get the value of $k_{init}$, again.

This polyalphabetic method achieves ciphering and deciphering of the vector's substrings. The algorithm is specified for a string of up to one hundred characters. If the length of a substring is not compatible with the expected length, the character $x$ isn't used as a filler at the encoding process. Space remains space. The same is true for the special characters. Also, lowercase letters remain lowercase and uppercase letters remain uppercase. 

\subsection{Example}
For the string: \textit{He surely likes Security and Bioethics, yes!} where $a=1$, $b=4$, and the key $k=3$, the solution will be:
\begin{center}
    \textit{Kh wyvjqd oloiw Xjfxumxc fsg Emsiymnfv, cix!} 
\end{center}
Since the key is equal to $3$, it will split the entire text into multiple groups of length $3$, with a possible exception of the last group of characters. 
Each character will be encoded, using a shift, specific to \textit{Caesar algorithm}, with $k$ positions to the right, at the encoding. This value increases after switching to the next character set. It will increase until the value $b+2$ is reached. Once it has reached, it will reset to the initial value of $k$. 

After splitting, the list will look like this:
[ “He ”, “sur”, “ely”, “ li”, ”kes”, “ Se”, “cur”, “ity”, “ an”, “d B”, “ioe”, “thi”, “cs,”, “ ye”, “s!” ]. 
The substring “He ” will become “Kh ” because it is shifted to the right by $k=3$ characters. For the next substring, $k$ will increase by $1$, so $k = 4$. Thus, "sur" becomes "wyv". In the next step, $k=5$; the string "ely" will become "jqd".
When the key $k$ increment to $6$, that means $k$ is greater than $b+1 = 5$. So, $k$ will be set to $3$. The process continues until the entire vector of substrings has been traversed. In the final step, each encoded substring will merge and the solution: 

\begin{center}
    \textit{Kh wyvjqd oloiw Xjfxumxc fsg Emsiymnfv, cix!} 
    
\end{center}
will be obtained. For the decoding part, it is important to mention that the process is similar to the encoding one. The principal difference is the mode of shifting the characters.

\subsection{Future directions }

Although the algorithm is simple to implement, it is uncertain whether it performs better than other methods in terms of execution time and storage space. With this meaning, a comparison of different ciphering and deciphering solutions with the proposed algorithm should be performed. 

What would be favorable to introduce in the future is a different variant of \textit{Caesar algorithm} for ciphering/deciphering. Namely, the encoding should be done by creating a list of English letters, both uppercase and lowercase. Therefore, a list of $52$ elements would be maintained, where uppercase letters would be entered first, as in ASCII encoding, in alphabetical order. The highest value that constant $a$ will be able to take will also be $47$, constant $b$ will be $50$ for this, and the value of $k$ will go up to $51$. Shifting will be performed using the traversal of the list of letters. 

Regarding the previous specifications, in the future variant, uppercase letters do not necessarily have to be uppercase, and lowercase letters do not have to be lowercase. The list will be traversed circularly, only once. This algorithm is simple to understand. But, it's not too hard for modern hackers to crack it. It could be compared, using different performance indicators, with other similar methods. Also, this method can be examined from the perspective of security. 

This paperwork is based on a scientific study and could be used as a basis for further research, starting with the Caesar cipher, Vigenère cipher, and Playfair algorithm, which can be improved.

How to keep sensitive data secret and secure is still evolving, but progress is being made \cite{IEEEhowto:a}. It would be useful to go in-depth with the necessary specifications and restrictions so that the developed, mixed algorithm is one with performing results. This would bring benefits both on the privacy and authorization side, as well as on the security of the data, and ciphers underlying the \textit{Latin Djokovic} algorithm implementation process.

\section{Conclusions }
In this section, we will specify our final observations and notes, and throughout the implication, we will make some conclusions. 

The three algorithms can be compared by analogy to the surfaces on which Novak Djokovic, the tennis player, is dominant. The algorithm may not be the best following each individual algorithm, but hopefully one day it will become powerful, and hard to exploit, combining different variations and processing into its nucleus existence.

Since we could not collect enough data to show the actual performance of the algorithm, we propose that further research and verification should be done on the topic and in the process of data transfer between the sender and receiver, as well as in the security scope of the database. 

Further studies can be carried out as a complement to this work, or we will continue with breakthrough research as specified. There are few works in this field, which modify and combine the different variations of the already existing methods. However, they can be useful if data security is considered.

In conclusion, it has been shown that ciphertext generated with Caesar and Vigenère cipher are prone to be easily broken using brute force, exhaustive search, and other methods because they lack diffusion and confusion. These are generated only with the modified hybrid of Caesar cipher and Vigenère cipher \cite{IEEEhowto:i}. However, this was at the beginning of the development of the algorithms. 

There is now a high percentage of diffusion in the algorithm that generates them, making the mixed cipher between Caesar and Vigenère very strong and rigid to crack. By specifying them, possibly the algorithm that includes the Playfair algorithm among them is an even better one. 


%

\appendices


\section*{Acknowledgment}
The author would like to express his thanks to the Department of Informatics, at the West University of Timișoara (UVT), Timiș, Romania, for the financial support at the registration stage for the conference, as well as some teaching staff who guided the author in terms of the structure of the text's attributes. The author wishes to express deep gratitude to the reviewers of this conference for their helpful suggestions that complement and punctuate the presentation of concepts in this paperwork. 

\ifCLASSOPTIONcaptionsoff
  \newpage
\fi



\begin{thebibliography}{1}

\bibitem{IEEEhowto:h}
Jain, A., et al. (2015). Enhancing the security of the Caesar cipher substitution method using a randomized approach for secure communication. arXiv:1512.05483. 

\bibitem{IEEEhowto:j}
Mishra, A. Enhancing security of Caesar cipher using different methods. International Journal of Research in Engineering and Technology. 

\bibitem{IEEEhowto:i}
Omolara, O. E., Oludare, A. I., \& Abdulahi, S. E. (2014). Developing a modified Hybrid Caesar cipher and Vigenère cipher for secure Data Communication. Computer Engineering and Intelligent Systems, 5(5), 34-46.

\bibitem{IEEEhowto:g}
Soofi, A. A., Riaz, I., \& Rasheed, U. (2016). An enhanced Vigenère cipher for data security. Int. J. Sci. Technol. Res, 5(3), 141-145. 

\bibitem{IEEEhowto:c}
Tunga, H., \& Mukherjee, S. (2012). A new modified Playfair algorithm based on frequency analysis. International Journal of Emerging Technology and Advanced Engineering, ISSN, 2250-2459.

\bibitem{IEEEhowto:b}
Patni, P. A poly-alphabetic approach to Caesar cipher algorithm. International Journal of Computer Science and Information Technologies, 4(6), 954-959.

\bibitem{IEEEhowto:a}
Sharma, V., Jalwa, S., Siddiqi, A. R., Gupta, I., \& Singh, A. K. (2021). A Lightweight Effective Randomized Caesar Cipher Algorithm for Security of Data. \emph{In Evolutionary Computing and Mobile Sustainable Networks}. Springer, Singapore. 

\bibitem{IEEEhowto:f}
Goyal, P., Sharma, G., \& Kushwah, S. S. (2015, December). A new modified Playfair algorithm using CBC. In 2015 International Conference on Computational Intelligence and Communication Networks (CICN) (pp. 1008-1012). IEEE. 

\bibitem{IEEEhowto:e}
Marzan, R. M., \& Sison, A. M. (2019, February). An enhanced key security of Playfair cipher algorithm. In Proceedings of the 2019 8th International Conference on Software and Computer Applications (pp. 457-461). 

\bibitem{IEEEhowto:d}
Dhenakaran, S. S., \& Ilayaraja, M. (2012). Extension of Playfair cipher using 16X16 matrix. International Journal of Computer Applications, 48(7).


\end{thebibliography}
%

%

\begin{IEEEbiographynophoto}
{Diogen Babuc} is a published author and informatician from Serbia. Babuc graduated from the Bioinformatics master's program as a valedictorian at the \textit{Faculty of Mathematics and Computer Science}, part of the \textit{West University of Timișoara}, Romania, in July 2023, and plans to start a Ph.D. program. Babuc has a degree in computer science from \textit{West University}. Since the beginning of 2022, he has been a teaching assistant at the \textit{Faculty of Mathematics and Computer Science} in Timișoara. Babuc is also a published informatician, who has participated in numerous scientific informatics meetings and conferences, and student and collegiate contests. 

In 2018, the publishing house \textit{Književna Opština Vršac} (Literary Municipality of Vršac) published Babuc's collection of poems and short stories, \textit{Kartagina}, in Serbian, based in its entirety on prominent but also hidden values from several periods, relying on the principle of general knowledge and action. In February 2023, the publishing house \textit{Nova Poetika} published his second book, \textit{Skladište}. His poems, stories, and translations have been published in the literary magazine of \textit{\textit{Književna Opština Vršac}}, and in \textit{Mladi dani} from Vrbas, Serbia.

\begin{figure}[h!]
    {\includegraphics[width=1in,height=1.25in,clip,keepaspectratio]{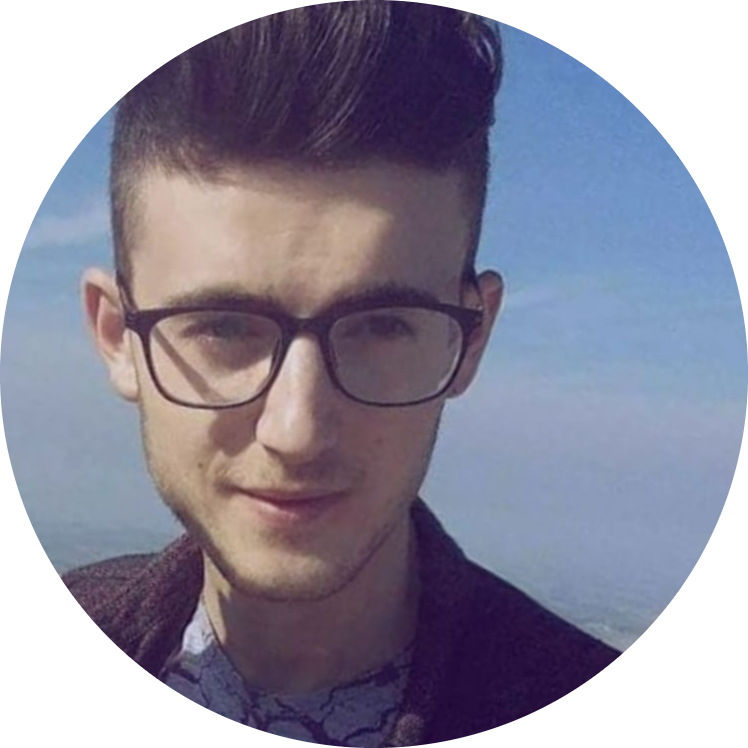}}{}
\end{figure} 

An IT organization, the \textit{Signal, Image Processing, and Machine Learning Team} (SIMT), accepted Babuc's web project with elements of artificial intelligence and machine learning called \textit{XAIBOT: Trilingual Erudite and Intelligent Chatbot} in 2020. A part of this project was integrated, a year later, into the \textit{Artificial Intelligence Hub} of the \textit{West University of Timișoara}.

He likes to sing and play the piano, as well as compose. He graduated from a junior music school. Babuc likes to play tennis and research data from the world of sports and regional geography. As a young computer scientist and writer, he strives to combine successes and experiences from the field of computer science and his own literary activities in his scientific-informatics and literary work, implementing them in diversified issues and trends. 
\\
\\
\textbf{Bibliography}: 
\\
Kartagina (Carthage), 10/2018, \textit{Književna Opština Vršac}, Vršac, Serbia.
\\ 
Skladište (The Storage), 02/2023, \textit{Nova Poetika}, Belgrade, Serbia. 

\end{IEEEbiographynophoto}




\end{document}